# Iron based superconductors: A brief overview

Haranath Ghosh and Smritijit Sen

Indus Synchrotron Utilization Division, Raja Ramanna Centre for Advanced Technology, Indore – 452013.

Fe-based superconductors were discovered in 2008. This discovery with $T_c$ values up to 56 $K$, generated a new belief in the field of superconductivity. Till its discovery, high temperature superconductivity in cuprates, created a prejudice that Cu-oxides are essential building blocks for a high temperature superconducting material. These Fe based superconductors do not contain Cu-O planes (some of the materials are even O free). It will be argued in this review, that these iron pnictide and chalcogenide (FePn/Ch) superconductors have Fe electrons at the Fermi surface together with an unusual Fermiology that can change rapidly with doping. This may lead to very different normal and superconducting state properties compared to those in standard electron-phonon coupled ''conventional'' superconductors. There are a large number of evidences showing that superconductivity, magnetism, orbital fluctuations are intimately related and coexist in these materials although the mechanism of superconductivity in these compounds is still unknown. The electronic specific heat, $2\Delta/k_B T_c$ ratio, phase diagrams, isotope effect, crystal structures and there correlation to $T_c$ from various available experimental data are main inputs of this review to show the above.

## INTRODUCTION

As iron has strong local magnetic moment, it is usually considered deleterious to superconductivity, the recent discovery of high temperature superconductivity at 26 K in LaFeAsO doped with F on the oxygen site in 2008 is one of the exceptions [1]. Other such examples of exceptions are $Th_7Fe_3$ ($T_c$ =1.8 K)[2], $U_6Fe$ ($T_c$ =3.9 K)[3], $Lu_2Fe_3Si_5$ ($T_c$ =6.1K)[4]. Furthermore, Fe itself under pressure is a superconductor with $T_c$=1.8 K at 20 Gpa [5]. Fe based superconductors do not have any copper-oxide layers which are thought to be essential ingredient for high $T_c$ cuprates. In fact some of the Fe-based superconductors with record $T_c$ of around 56 K, neither contains $Cu$ nor $O$, e.g., $Gd_{0.8}Th_{0.2}FeAsO$, $Sr_{0.5}Sm_{0.5}FeAsF$ and $Ca_{0.4}Nd_{0.6}FeAsF$ [6]. Fe based superconductors also known as Fe-pnictides (''$FePn$'' where $Pn$ is $As$ or $P$) have extended its family to include iron chalcogenides (''$FeCh$'' where $Ch$ includes $S$, $Se$ and $Te$). We shall firstly discuss about structural aspects and its correlation to $T_c$ in these compounds below. Strikingly, more than 3000 papers have already been published about Fe-based superconductors. These suggest that the superconductivity may be related to the coexistent magnetism, orbital fluctuation and primarily not due to phonon. The properties of the FePn/Ch superconductors are fundamentally different both from those of a conventional electron-phonon coupled superconductor and from those of the high $T_c$ cuprates. It is accepted (by now) that the superconducting pairing symmetry in high $T_c$ cuprates is $d_{x^2-y^2}$ kind corresponding to $l = 2$ orbital angular momentum. Such type of unconventional order parameter allows finite electronic excitations remaining even at $T \to 0$. This is in contrast to a clean conventional superconductor, where the electronic excitations are suppressed (exponentially) below $T_c$ by opening a gap in the electronic spectrum. Such existence of finite electronic excitations well below $T_c$ is also found from numerous experimental studies in Fe-based superconductors. While superconducting order parameter symmetry and the mechanism responsible are still unknown, it is apparently not conventional $s$ wave symmetry. Neutron scattering measurements provide convincing evidence for a sign change in the superconducting energy gap on different parts of the Fermi surface in a number of compounds. Thus Fe-based superconductors are not only unconventional but also different from high-$T_c$ cuprates. We shall argue that Fe-based superconductors are *different* from others in a number of ways; (a) coupling of magnetism, superconductivity and orbital degrees of freedom --- possible connection of structural transition to orbital degree than lattice. In a number of cases structural and magnetic transition temperatures are identical. (b) The specific heat jump scales very differently than other conventional as well as unconventional superconductors. (c) They have very different Fermi-ology made of Fe-$d$ electrons than other classes of superconductors which changes drastically with electron or hole doping. (d) Although numerous theoretical/experimental data suggest to multiband Fermi surface and multi-gap, there are strong spectroscopic evidences of two –gaps (i) one large and (ii) small gaps giving rise to two sets of $2\Delta/k_B T_c$ ratios, averaging about 7 and 3 respectively. (d) Unlike conventional superconductors Fe-based superconductors do not exhibit any $O$ istope effect but Fe istope effect instead. (e) The spin susceptibility in a large number of compounds produces linear temperature dependencies. (f) Spin resonance energy which is also observed in high $T_c$ cuprates, scales with $T_c$ linearly. We shall follow in the rest of the article providing support to the above directions from available theoretical and experimental studies.

**STRUCTURAL ASPECTS AND CORRELATION TO $T_c$**

There are a large number of Fe-based compounds (see Fig. 1) which may broadly be classified into two categories viz, Fe-pnictides ($FePn$) and Fe-

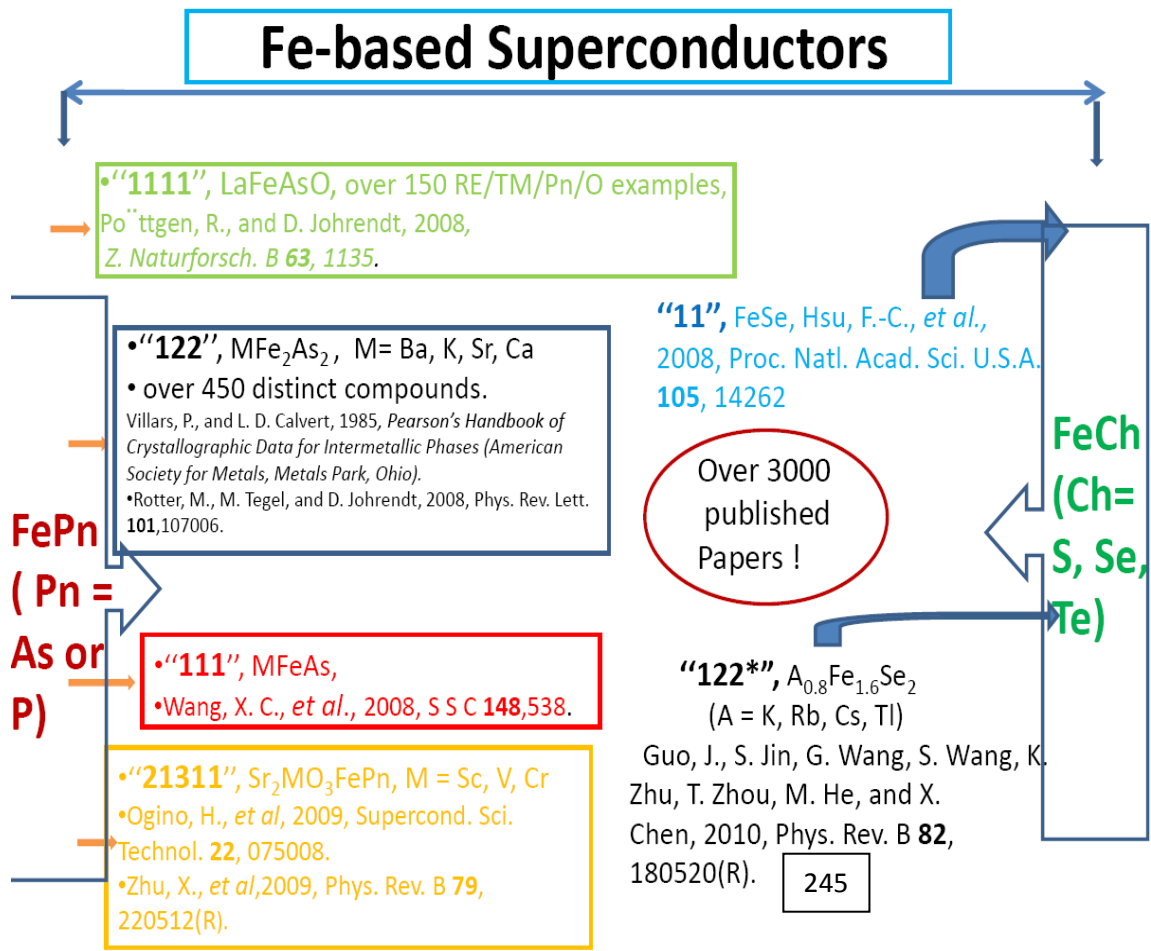

**Figure1**.*Families of Fe-based superconductors along with discoverers and associated references.*

chalcogenides ($FeCh$). The former can further be classified into four categories whereas the later into two, giving rise to a total of six categories of Fe-based materials as presented in Fig.1. First category of Fe-based materials, known as 1111 compounds, is made up of the combinations of elements, rare-earth ($La, Sm, Gd$ .... etc.)-transition metal ($Fe, Co, Ni$....)-pnictogen ($As, P$)-oxygen; more than 150 such compounds are possible [7] out of which several are superconductors. In Fig. 2 we present crystal structure of $LaOFeAs$ which is representative of 1111 family. They have the tetragonal, $tP8$ (''t'' means tetragonal, ''$P$'' means ''primitive''or no atoms in either the body or face centre, 8 atoms per unit cell) $ZrCuSiAs$ (prototypical compound) structure with 2D layers of $FeAs$. They belong to space group $P4/nmm$ with space group number 129 (tetragonal). These materials undergo tetragonal to orthorhombic structural transition and have space group $Cmma$ or $P112/n$ or $P2/c$ in the orthorhombic phase.

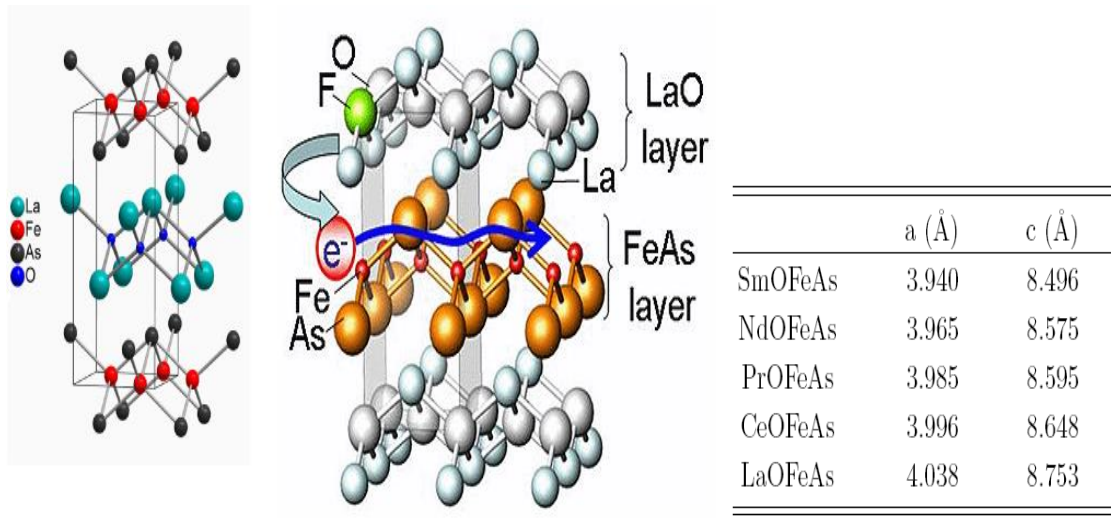

**Figure2**. *Crystal structures of LaOFeAs compound a representative of 1111 family with lattice parameters on the right ($c/a$ ~2.16).*

The most studied and probably the largest numbers of superconductors are in the 122 family which are made up of combinations of metallic alkaline earth ($Ba, K, Ca, Sr, …$)-transition metal (Fe, Co, Ni…)-pnictogen (As,P). More than 450 compounds are possible within these combinations (see the reference in Fig.1). Crystal structure of $BaFe_2As_2$ is presented in Fig. 3. These materials do not have any oxygen; hence instead of $LaO$ layer that

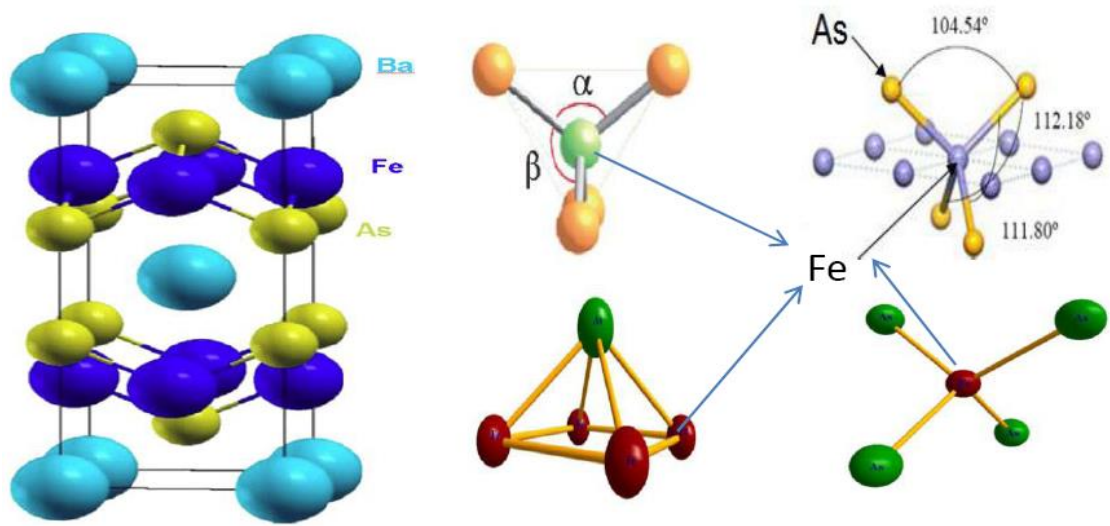

**Figure3.** *Crystal structure of $BaFe_2As_2$ and others in the 122 family. $FeAs_4$ tetrahedron in different orientations, various $As-Fe-As$ bond angle, $Fe-As-Fe$ bond angle, pnictogen height from the Fe-plane is shown. These parameters strongly influence superconducting properties.*

separates FeAs layers in 1111 compounds are replaced by a single $Ba$ atom. Similar to the 1111 compounds they are also in tetragonal structure (with 2D FeAs planes common to all the Fe-based materials) with $tI10$, standard $ThCr_2Si_2$ structure. Here 't' stands for tetragonal, ''$I$ 10'' refer to the fact that there is an atom at the centre of the 10 atom unit cell. They have the same structure as that of the first discovered heavy fermion superconductor, $CeCu_2Si_2$ [8]. $BaFe_2As_2$ has $T_c$ of 38 K on potassium (K) doping and was discovered by Rotter *et al.*,[9].

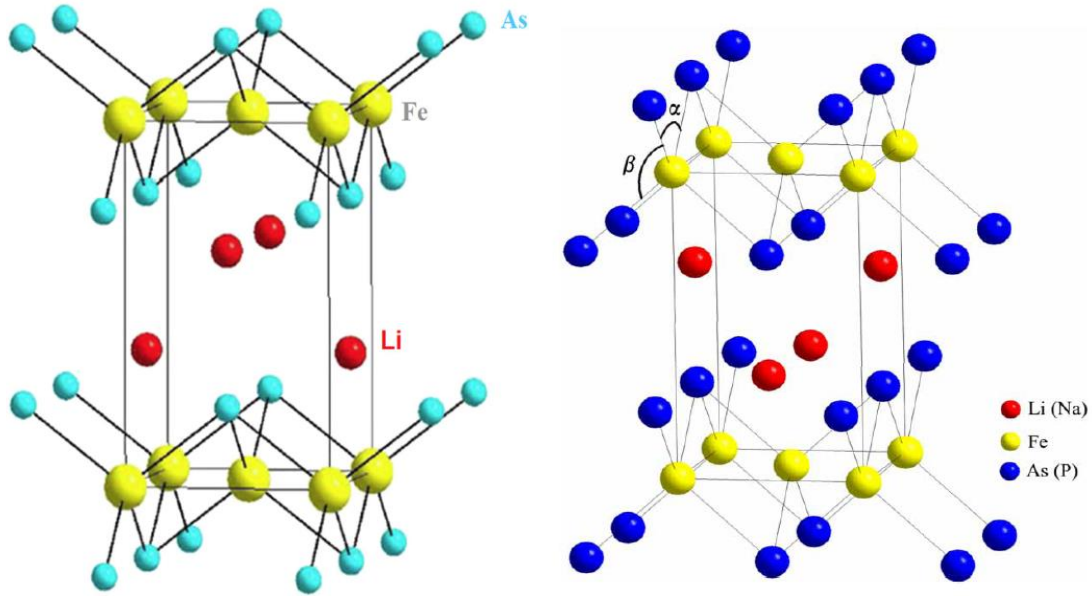

**Figure4**. *Typical crystal structures of 111 LiFeAs. The left Figure is taken from [10] whereas the right figure from [11].*

The "111"-type iron-based superconductors crystallize into a $Cu_2Sb$-type structure with $P4/nmm$ symmetry, as shown in Figure 4. In the layered structure, iron pnictide and a $Li$ or $Na$ layer are stacked alternately. Fe atoms are in a four-fold coordination, forming a $FeAs\,(P)_4$ tetrahedron (see right hand side figure of Fig. 3). Table 1 shows the parameters of these crystal structures. These compounds have varied $c/a$ ratio compared to other classes discussed above, *e.g*, $c/a = 1.684$ for $LiFeAs$, $=1.782$ for $NaFeAs$ and 1.634 for $LiFeP$ respectively. The crystal lattice parameters of $LiFeP$ are the smallest among the three "111" compounds. The bond angle for a regular tetrahedron (cf. Fig.3) is about 109.47∘. From Table 1, we can see that for both $NaFeAs$ and $LiFeP$, the $FeAs(P)_4$ tetrahedron approaches the regular one, while the $FeAs_4$ tetrahedron for $LiFeAs$ shows elongation along the *c*-axis, which leads to a much smaller $As$–$Fe$–$As$ (two-fold) angle and a larger anion height from $As$ atom to $Fe$ plane than that of $NaFeAs$. Correlations of various angles of $As-Fe-As$ tetrahedron, anion height from As atom to Fe plane (pnictide height) with $T_c$ is discussed below.

|  | $a$ (Å) | $c$ (Å) | $\alpha^a$ | $\beta^a$ | Anion height (Å) |
|---|---|---|---|---|---|
| LiFeAs | 3.776 | 6.358 | 102.8 | 112.9 | 1.51 |
| NaFeAs | 3.949 | 7.039 | 108.3 | 110.1 | 1.43 |
| LiFeP | 3.692 | 6.031 | 108.6 | 109.9 | 1.33 |

Note: $^a\alpha$ and $\beta$ denote the two-fold and four-fold bond angle of the FeAs(P)$_4$ tetrahedron, respectively.

**Table 1**. *Crystal structure parameters for "111"-type superconductors under ambient pressure and room temperature.*

The rest family of $FePn$ materials, 21311 (sometimes also called the 42622) is shown in Fig. 5. The structure in Fig. 5 can be visualized as layers of 122, $SrFe_2P_2$ alternating with perovskite $Sr_3Sc_2O_6$ layers (so $SrFe_2P_2$ together with $Sr_3Sc_2O_6$ making it to $Sr_4Sc_2O_6Fe_2O_2$ and hence the name 42622). Intercalation of further layers of atoms between the $FeAs$ layers to try to increase $T_c$ by expanding the c axis has so far [12] resulted in $T_c$ s up to 47 K. This family has same space group, $P4/nmm$, space group number 129. The symbol $nmm$ means symmetric about mirror planes perpendicular to the two equal tetragonal axes ($a$ and $b$) and that for the third, unequal tetragonal axis ($c$ -axis) the symmetry operations that bring the crystal back to itself are called glide plane symmetry, where the $n$ glide involves reflecting about a mirror plane parallel to the $c$-axis followed by a translation along 1/2 of the face diagonal.

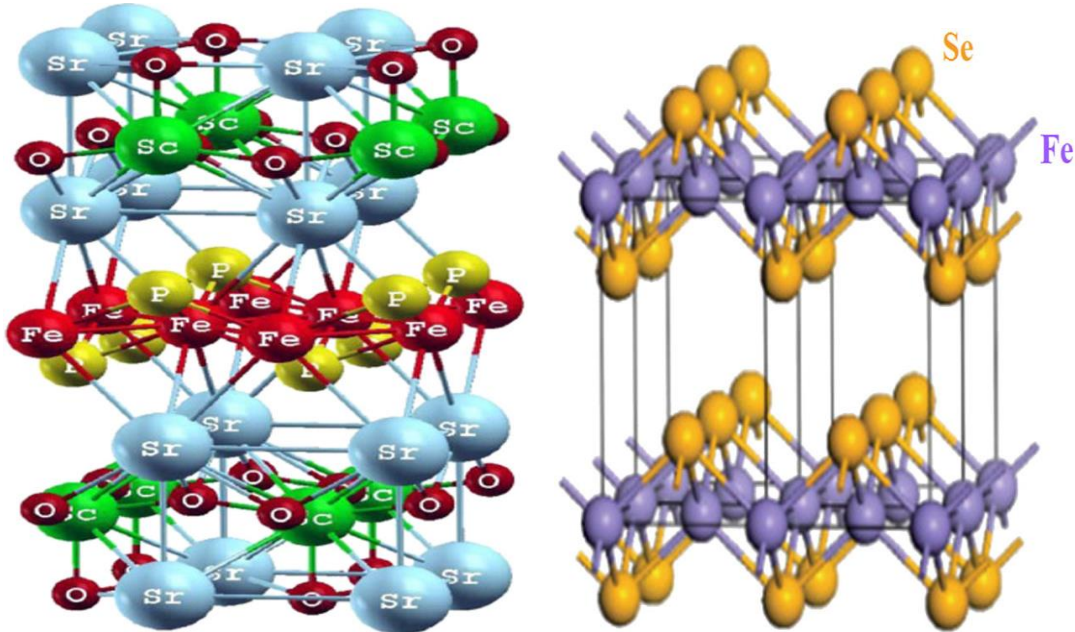

**Figure 6.** *Lattice structure of $Sr_2ScO_3FeP$ taken from [12] (left side). Lattice structures of 11 structure (first category on the right hand side of Fig.1) FeSe taken from [13]. For details see text.*

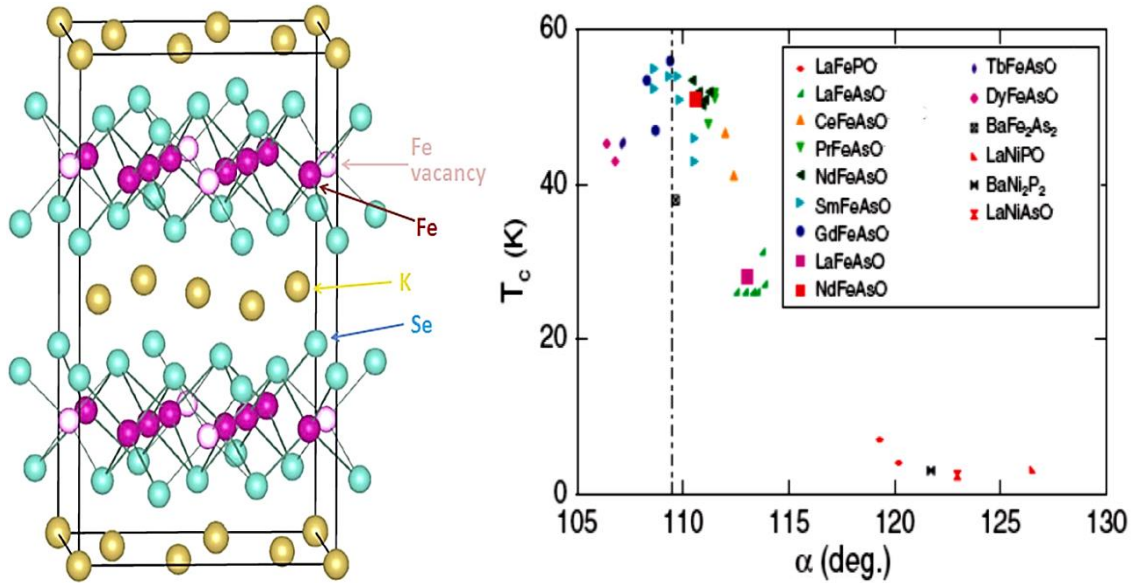

**Figure 7.** *(left) Typical crystal structure of 122* materials, has reduced symmetry due to ordered Fe vacancies (compare with Fig. 3). (Right) Variation of superconducting transition temperature ($T_c$) with angle $As - Fe - As$ defined in Fig.3 and Table-I, taken from [13].*

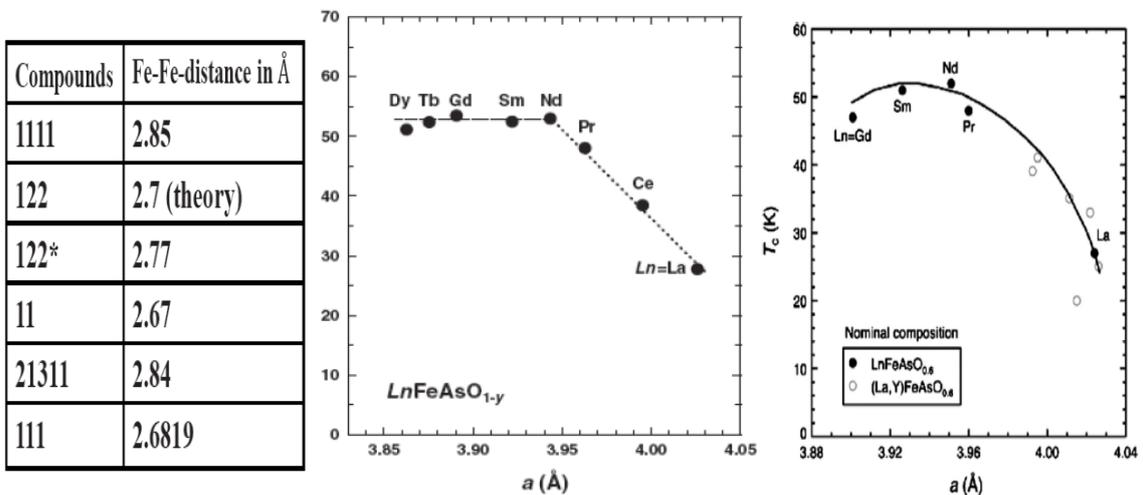

**Figure.8** *(left) Typical $Fe - Fe$ distances in various Fe-based compounds. They are small indicating that the Fermi surface would principally be made of $Fe - 3d$. The right hand side figures present variation of $T_c$ with lattice constants. The extreme right figure presents the same as the middle figure but includes results for Y replacing La in 1111 compounds.*

The most recent $FePn/Ch$ structure discovered (Fig. 7, left) with superconductivity ($T_c = 32\ K$) is an ordered defect alteration of the 122 $BaFe_2As_2$ structure (called the ''122*'' structure herein), written as $A_{0.8}Fe_{1.6}Se_2$ or sometimes $A_xFe_{2-y}Se_2$ ($A = K, Rb, Cs, Tl$), where the ordered arrangement of $Fe$ vacancies below structural transition on the inequivalent Fe sites (in the ideal case $Fe2$ sites are fully occupied, $Fe1$ sites are fully unoccupied) has important influence [15,16] on the measured properties, including superconductivity. These materials have exceptionally high Neel temperature ($T_N > 500\ K$) and magnetic moment (~3 $\mu_B$/Fe-atom). This family is often called "245" because of its parent compound $A_{0.8}Fe_{1.6}Se_2 \equiv A_2Fe_4Se_5$. This structure may alternatively be viewed as $FeSe$ intercalated with $K, Rb, Cs, Tl$, or combinations thereof. The unit cell for the tetragonal 122* ordered defect structure is larger than that for the tetragonal 122 by $\sqrt{5} \times \sqrt{5} \times 1$ in the $a, b$, and $c$ axis directions, respectively; see [15,16] for further diagrams. The 122* (or the 245) structure has the reduced $I4/m$ symmetry (space group 87) below the defect ordering transition $T_S$ (vs $I4/mmm$ of the 122 structure at higher temperatures) the ordered-defects cause loss of mirror plane symmetries in the $x$ and $y$ directions of the 122 structure when the $Fe1$ sites are empty (compare Fig. 3 with Fig. 7). Therefore, four of the six structures belong to $P4/nmm$, space group number 129. The other two $MFe_2As_2$ (122) and (122*/245) have $I4/mmm$ (space group number 139) and $I4/m$ (space group number 87) respectively.

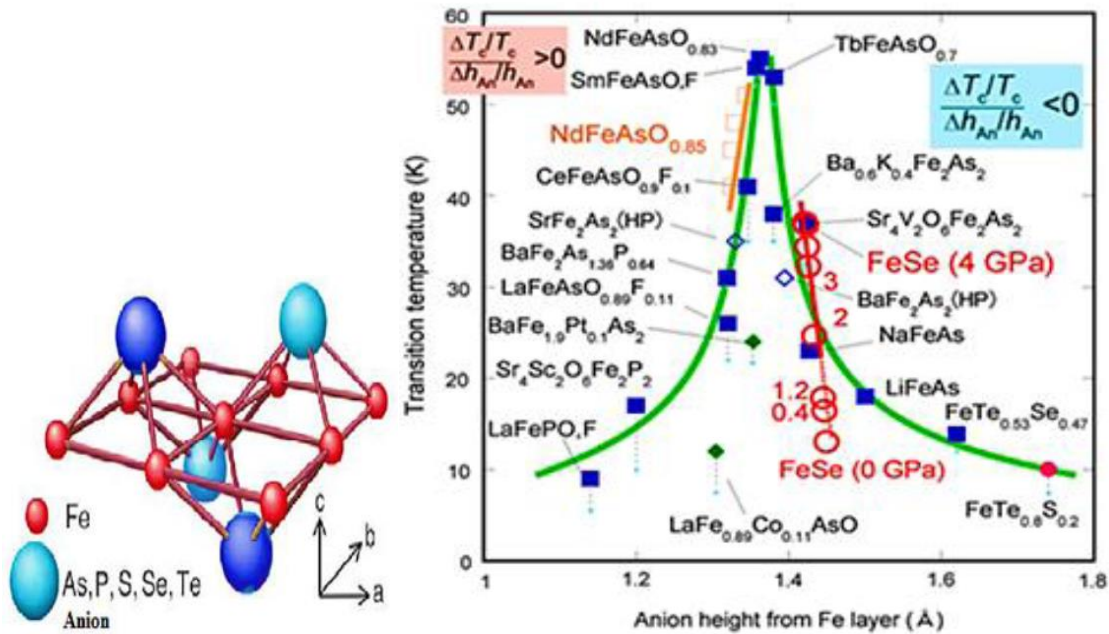

**Figure** 9. *Variation of $T_c$ as a function of anion height of all Fe-based materials (some of the results presented are under pressure see [17] for details). When the pnictide height (see the left figure) approaches 1.4 Å highest $T_c$ is obtained.*

Fig. 9 shows a symmetric curve with a peak around 1.38Å. The zero-resistivity temperatures at ambient pressure are indicated by small light-blue circle. Filled diamonds indicate the data at ambient pressure. This dependency of anion height with $T_c$ is an important correlation with structural aspects as it may help search for new Fe-based superconductors with higher $T_c$ and also shed light on the mechanism of Fe-based superconductors. We shall come back to the discussion of figure 9 when presenting isotope effect in these materials. Correlation of superconducting $T_c$ with various structural parameters are demonstrated in Fig. 7 (right), Fig. 8 and 9 respectively. It is clear that irrespective of mechanism of superconductivity highest $T_c$ are attained only for α~110° (Fig.7), lattice parameter 3.96 Å (Fig.8) and anion height 1.38 Å (Fig.9) respectively indicating close correlation between structural aspects and superconductivity in these materials.

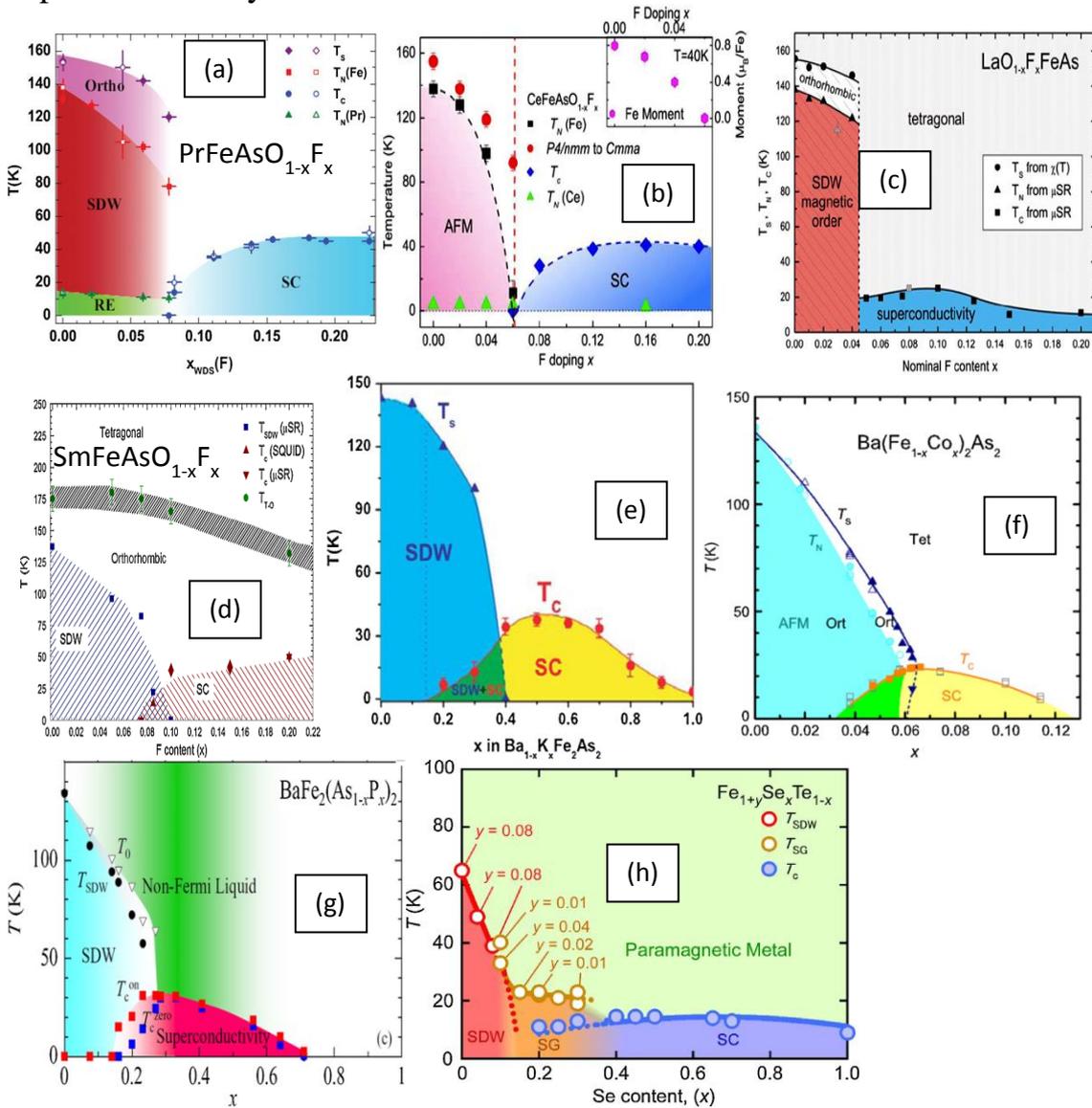

**Figure 10.** *Phase diagrams of Fe-based materials: 1111 (a-d), 122 (e-g), 245(h) (taken from [18,19,20]).*

**PHASE DIAGRAMS**

Phase diagrams of different classes of Fe-based superconductors are presented in Fig. 10. Phase diagrams (a) to (d) correspond to various 1111 materials. Figures 10 (e to f) correspond to various 122 materials whereas 10 (h) is for 245 materials for details of these phase diagrams see various references in [18,19,20]. Out of all these, phase diagrams of 122 materials are well established due to availability of good quality samples. These materials (122) are also most extensively studied and very large number of materials exists in this family, as one can dope on any of the three sites M (e), Fe (f) or As (g). These phase diagrams are obtained through resistivity, spin susceptibility, $\mu$ SR measurements etc.; see for example various references in [18, 19, 20]. These materials have well established magnetism and superconductivity coexistence. The magnetism is caused by spin density wave (SDW) type transition seen in resistivity, spin susceptibility and neutron scattering measurements. It is worth noting that there exists structural transition ($T_S$) together with the SDW transition and for M-site doped materials they are identical or nearly identical whereas for Fe-site doping (with Co) it differs as doping increases. The trend is also similar in case of As-site doped materials. These constitute the central theoretical and experimental studies of Fe-based materials. In 1111 materials while SDW and structural (tetragonal to orthorhombic) transitions are present but there are no significant coexistence between the SDW and superconductivity (SC) is observed. While structural transition is always slightly higher than the SDW transition, phase diagram for the Sm-compound came as a surprise, the $T_S$ is way above SDW transition, casting doubt about earlier understanding that the structural, magnetic and superconducting transitions are intertwined. Below is a detailed comparison of $T_S$ and $T_{SDW}$ among 1111 and 122 alongwith other Fe-based compounds. It is noticeable that for 122, 11 un-doped compounds they are identical whereas the difference between them decreases as sample quality improves for 1111 compounds (2$^{nd}$ rows).

| Material | TS(K) | TSDW(K) | Reference |
|---|---|---|---|
| LaFeAsO | 158 | 134 | Luetkens et al.,[18] |
| PrFeAsO | 154 | 135 | Rotundu et al.,[18] |
| CeFeAsO | 155 | 140 | Zhao et al., [18] [20] |
|  | **151** | **145** |  |
| NdFeAsO | 150 | 141 | Qiu et al., [21], Chen et al., [21] |
|  | 143 | 137 | Tian et al. [21]. |
| SmFeAsO | 175 | 135 | Martinelli et al. [18], Drew et al., [18]. |
| GdFeAsO | 135 | ??? | C. Wang et al., [22]. |
| SrFeAsF | 180 | 133 | Xiao et al.,[22] |
| CaFeAsF | 134 | 114 | Xiao et al.,[22]. |
| BaFe$_2$As$_2$ | *142* | *142* | Huang et al.,[23] |
| SrFe$_2$As$_2$ | *205* | *205* | Krellner et al. [23] |
| CaFe$_2$As$_2$ | *171* | *171* | Ronning et al.,[23] |
| EuFe$_2$As$_2$ | *190* | *190* | Tegel et al., [23] |
| FeTe | *72* | *72* | Mater. Res. Bull. **10**, 169; Phys. Rev. B **81**, 094115. |
| FeSe | *90* | ---- | Phys.Rev. Lett. **103**, 057002. |
| LiFeAs | ------ | ------ | X.C. Wang et al., SSC **148**, 538. |
| $Na_{1-\delta}FeAs$ | *50* | *40* | Phys. Rev. B **80**, 020504(R). |
| $K_{0.8}Fe_{2-y}Se_2$ | *578/551* | *559/540* | arXiv:1102.3674 ;EPL **94**, 27008 |
| $Rb_{0.8}Fe_{\{2-y\}}Se_2$ | *540* | *534* | Liu et al. (2011) EPL **94**, 27008. |
| $Cs_{0.8}Fe_{\{2-y\}}Se_2$ | *525* | *504* | Liu et al. (2011) EPL **94**, 27008. |
| $Sr_2VO_3FeAs$ | *Sample sensitive* | 155 K | Cao et al., Phys. Rev. B **82**, 104518. |

Landau theory states that two simultaneous phase transitions that interact with each other (*i.e.*, are not simultaneous due to coincidence) and break different symmetries result in a first order transition. Wilson *et al.* (2009), in their neutron scattering experiments on a high quality single crystal of $BaFe_2As_2$, found that both the structural and magnetic transitions at 136 K are second order. Tegel *et al.*, (2008) [23] argued from their measurements of the lattice order parameter [ $P = \frac{a-b}{a+b}$, where $a$ and $b$ are the orthorhombic axes' lengths] in $M = Sr$ ($T_S = 203\ K$) and $Eu$ ($T_S = 190\ K$) that, despite their measured cell volume discontinuity at $T_S$ in $SrFe_2As_2$, all of the $MFe_2As_2$ starting compounds undergo in fact second order structural phase transitions. In light of the prediction of Landau theory, then either the simultaneity of $T_S$ and $T_{SDW}$ are coincidental or there should be some higher temperature precursor of one of the transitions that breaks that transition's symmetry at a higher temperature. Ghosh *et al.*, (2012) [24] showed that if $T_S$ is due to orbital modulations and $T_{SDW}$ is interorbital in origin due to nesting of Fermi Surface then they can occur simultaneously. Before we conclude to this section we shall describe isotope effect in these materials which is known as diagnostics of electron-phonon mediated BCS superconductivity.

**ISOTOPE EFFECT**

Co-existence of superconductivity and magnetism in FePn superconductors, show a strong sensitivity to the crystal lattice, suggesting the possibility of unconventional electron–phonon coupling. Liu *et al.*, reported the effect of oxygen and iron isotope exchange on $T_c$ and the spin-density wave (SDW) transition temperature ($T_{SDW}$) in the $SmFeAsO_{1-x}F_x$ and $Ba_{1-x}K_xFe_2As_2$ systems [25] (see fig. 11). They also conclude that the impact of oxygen isotope effect on $T_c$ and $T_{SDW}$ is very small whereas iron isotope effect exponent $\alpha_c$ is about 0.35 (for full isotope effect $\alpha_c =$ 0.5). Another important thing is that exchange of iron isotope affect $T_c$ and $T_{SDW}$ same way. So electron-phonon interaction plays a significant role in these FePn superconductors but strong magnon-phonon coupling creels conventional electron-phonon coupling and makes it unconventional.

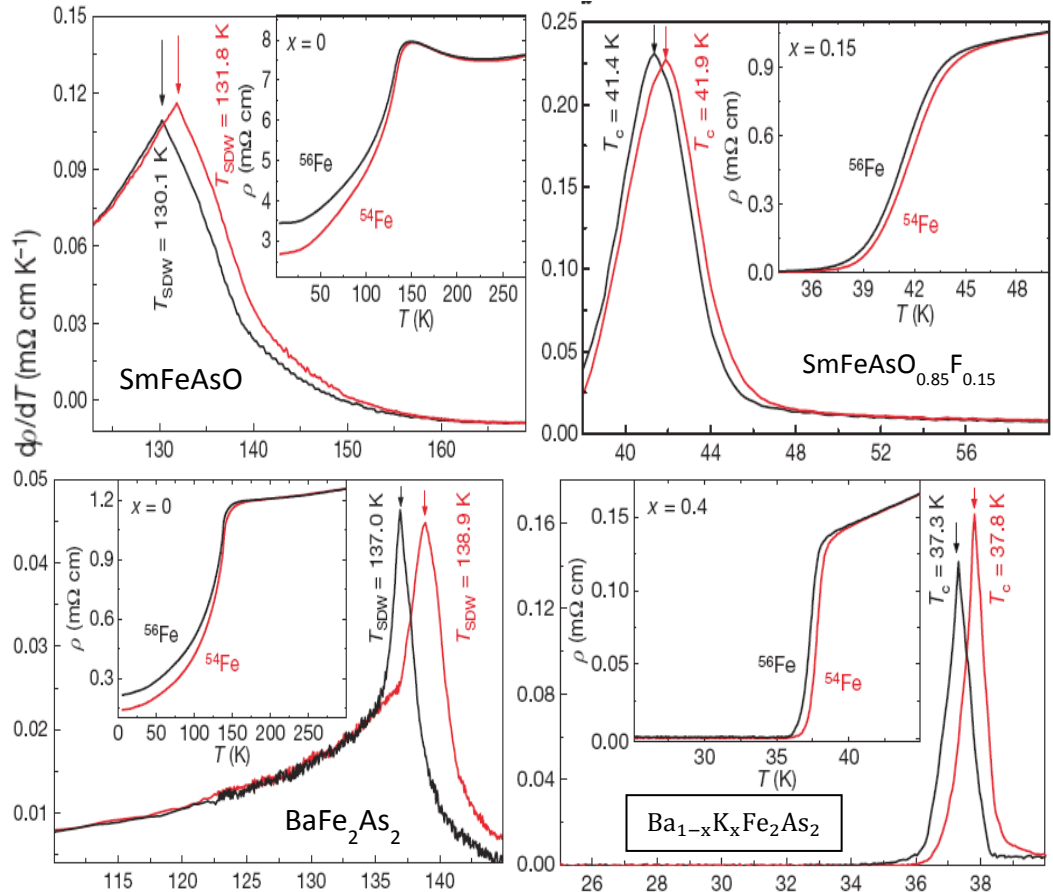

**Figure. 11** *Temperature dependence of $\rho(T)$ and $\frac{d\rho(T)}{dT}$ with Fe istope $^{56}Fe$ & $^{54}Fe$ substitutions (reproduced from ref [25]).*

Isotope effect in Fe-based superconductors is currently controversial. It has been found that value of isotope effect exponent $\alpha$ for Fe-based superconductors may be negative as well as positive and in few cases it exceeds BCS value $\alpha_{BCS} = 0.5$. Associated with Fe-isotope substitution structural changes occur simultaneously (causing changes in $\alpha$ defined in Fig. 3 and pnictogen height defined in Fig. 9 and hence the associated change in $T_c$). As a result, Fe isotope exponent contains two parts, $\alpha_{Fe} = \alpha_{Fe}^{int} + \alpha_{Fe}^{str}$; one is related to structural changes and other one is intrinsic. A table below provides estimation of various parameters like $T_c$ of natural Fe isotope $\alpha_{Fe}, \alpha_{Fe}^{int}, \alpha_{Fe}^{str}$, relative shift of the c-axis lattice constant due to isotope substitution [17, 26-29].

| Sample | $T_c(^{nat}Fe)$ (K) | $\alpha_{Fe}$ | c-axis($^{light}$Fe) (Å) | c-axis($^{heavy}$Fe) (Å) | $\Delta c/c$ | $\alpha_{Fe}^{str}$ | $\alpha_{Fe}^{int}$ | Reference |
|---|---|---|---|---|---|---|---|---|
| FeSe$_{1-x}$ | 8.21(4) | 0.81(15) | 5.48683(9) | 5.48787(9) | >0 | ≃0.4 | ≃0.4 | R.Khasanov et.al 2010 |
| Ba$_{0.6}$K$_{0.4}$Fe$_2$As$_2$ | 37.30(2) | 0.37(3) | 13.289(7) | 13.288(7) | ~0 | ~0 | ~0.35 | R.H. Liu et.al 2009 |
| Ba$_{0.6}$K$_{0.4}$Fe$_2$As$_2$ | 37.78(2) | −0.18(3) | 13.313(1) | 13.310(1) | <0 | ~−0.5 | − | P.M. Shirage et.al 2009 |
| SmFeAsO$_{0.85}$F$_{0.15}$ | 41.40(2) | 0.34(3) | 8.490(2) | 8.491(2) | ~0 | ~0 | ~0.35 | R.H. Liu et.al 2009 |
| SmFeAsO$_{1-y}$ | 54.02(13) | −0.024(15) | 8.4428(8) | 8.4440(8) | ≳0 | <0 | − | P.M. Shirage et.al 2010 |

**Resistivity & Susceptibility:** Since the phase diagrams are obtained mostly from resistivity and susceptibility measurements we provide a very brief review of the same in this subsection. Resistivity and susceptibility measurement provide clue about the progression of these parameters with doping via various anomalies (structural, magnetic) in these parameters. In addition, the residual resistivity ratio $\rho/\rho_0 (RRR)$ is the indicator of sample quality as impurity scattering increases the residual resistivity $\rho_0$. Temperature dependence of resistivity also provides important perception about the nature of the normal state as well as superconductivity of these FePn/Ch superconductors. For example, in case of high $T_c$ cuprates resistivity linearly vary with T indicating very unconventional normal state. Typical behaviour of resistivity is shown in Fig.12. In general, FePn/Ch superconductors have metallic behaviour $d\rho/dT > 0$. For undoped and Co doped $SrFe_2As_2$ and $FeSe$, the temperature dependence of resistivity is presented in Fig. 12. But evidence of non-metallic behaviour is reported by Mizuguchi *et al.*,(2009) in FeSe$_{1-x}$Te$_x$, $x > 0.25$. It was believed that, all recently discovered A (= K, Rb etc.) intercalated iron selenide superconductors share the common crystalline and magnetic structure, which are very different from previous families of Fe-based superconductors [32] and constitute a distinct new 245 family. But more recently, $^{77}Se$ and $^{87}Rb$ nuclear magnetic resonance (NMR) study in the iron-Selenide $Rb_{0.74}Fe_{1.6}Se_2$ reveal clearly distinct spectra originating from a majority antiferromagnetic (AF) and a minority metallic superconducting (SC) phase. Their findings show that the SC phase in the bulk single crystals of 245 iron-selenides has the composition $Rb_{0.3(1)}Fe_2Se_2$, *i.e.*, FeSe layers with a doping of 0.15 electrons per Fe. At such a doping level, most Fe-pnictide compounds are known to exhibit superconductivity. The coexistence of superconductivity and strong local moment antiferromagnetism in the 245 iron-selenide was taken as evidence in favor of a strongly correlated Mott-insulator scenario and against a more weak-coupling approach based on an itinerant nesting picture. But it is demonstrated in ref. [32] that the AF and SC phases segregate, and that the SC phase in the 245 compounds has nothing very specific, but is merely an electron doped iron-selenide layer without any Fe vacancies and with no local Fe moments. Thus, the Mott picture cannot be argued to explain superconductivity in the 245 iron selenide family. These compounds remain however original and interesting as they display a natural hetero-structure where superconducting layers alternate with a peculiar AF state containing ordered Fe vacancies.

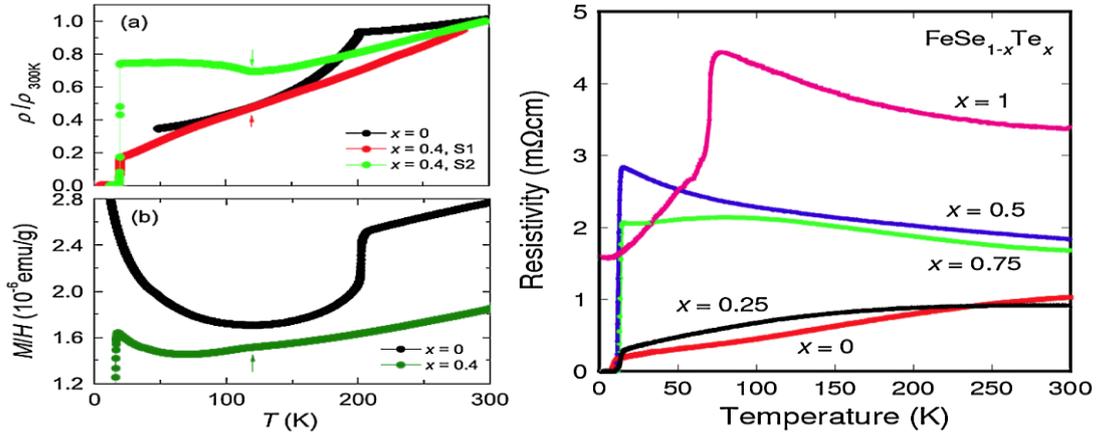

**Figure 12.** *(left)Temperature dependence of resistivity (a) and susceptibility (b) of $SrFe_{2-x}Co_xAs_2$ single crystals (taken from [30]). (**Right**) Temperature dependence of resistivity (taken from [31]) of $FeSe_{1-x}Te_x$ (polycrystalline).*

In FePn/Ch magnetic susceptibility $\chi$, displays large anomaly at $T_{SDW}$ see Fig.13, showing $\chi \sim T$ above $T_{SDW}$ indicating magnetic fluctuations in the normal state. Earlier studies suggested that, the structural transition is due to a fluctuating magnetic state without long range order [33]. More recent research reveals that the structural transition (breaking of tetragonal $a - b$ axis symmetry) is due to nematic magnetic fluctuations. Apart from these a number of other theories are also available in the literature, which put forward the idea that orbital ordering plays an important role in structural transition. Five $d$-orbitals of Fe among which two in the directions that are asymmetric in $xy-$plane and could play an important role in the structural transition (follow sections below).

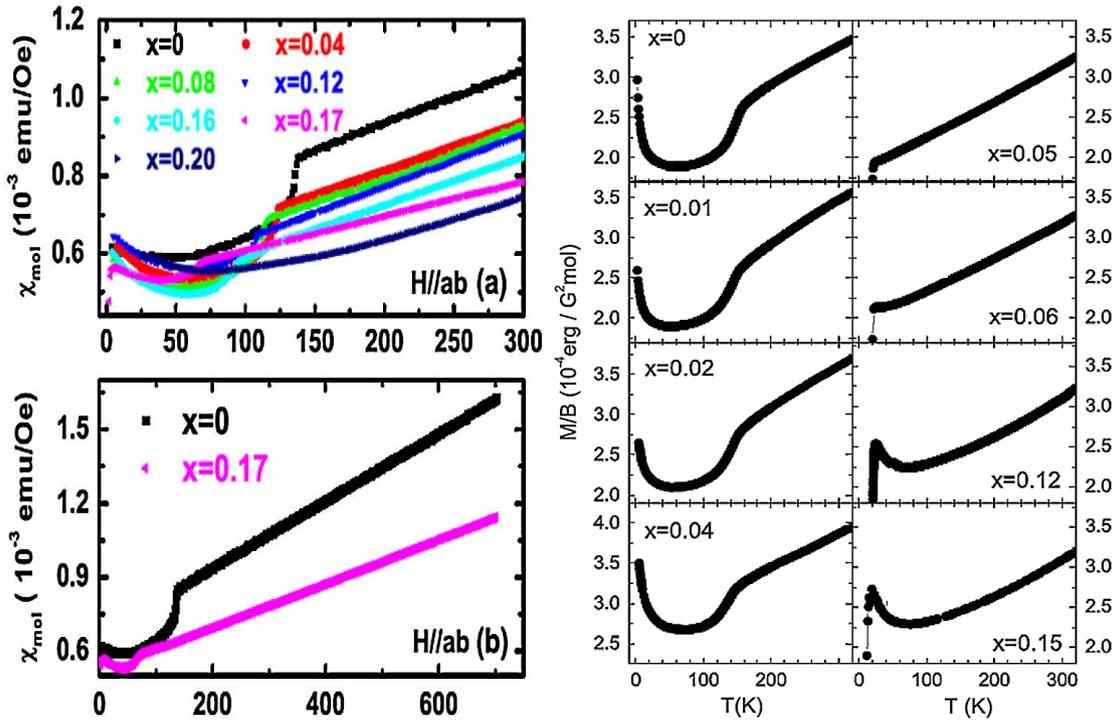

Figure 13. (left) Magnetic susceptibility ($\chi(T)$) for $BaFe_{2-x}Co_xAs_2$. The linearity of $\chi(T)$ with $T$ disappears abruptly for $x = 0.20$ (courtesy X. F. Wang et al., 2009). $\chi(T)$ for $LaFeAsO_{1-x}F_x$, $0 < x < 15$. The large anomalies at $T_{SDW}$ up to $x = 0.04$ (taken from Klingeler et al., 2010).

## Specific Heat:

Specific heat studies of FePn/Ch superconductors reveal informations regarding higher temperature transitions, like structural and magnetic (SDW) transitions along with SC transition. Because of higher $T_c$s, phonon contribution to specific heat (C) is too large to find specific heat jump accurately. This may be achieved in two ways (a) by applying large enough magnetic field successively to cause suppression in SC and extrapolating it to the normal state, (b) by using a neighbouring composition (for example, replacing Fe by Co as they have almost same molar mass) that is not superconducting. Such contribution of the normal state then may be deducted to get the electronic contribution to the specific heat below $T_c$. In the first case (a) C/T extrapolated to T=0 from normal state data gives

Sommerfeld constant $\gamma_n = \lim_{T\to 0} C_{normal}/T$, which is proportional to the renormalized bare electronic density of states at the Fermi level N(0). $\gamma_n \sim (1+\lambda)N(0)$, where $\lambda$ is contribution due to electron-phonon as well as electron-electron interactions present in a material. It is a useful parameter that can be related to band structure calculations of N(0) & de Has van Alphen measurement of effective masses of various Fermi surface orbits ($\gamma_n \propto m^*$) ($m^*$ being effective mass). Following methods (a,b) typical electronic specific heat data are presented in Fig. 14. The difficulty in measuring the specific heat jump may be appreciated from right side Fig.

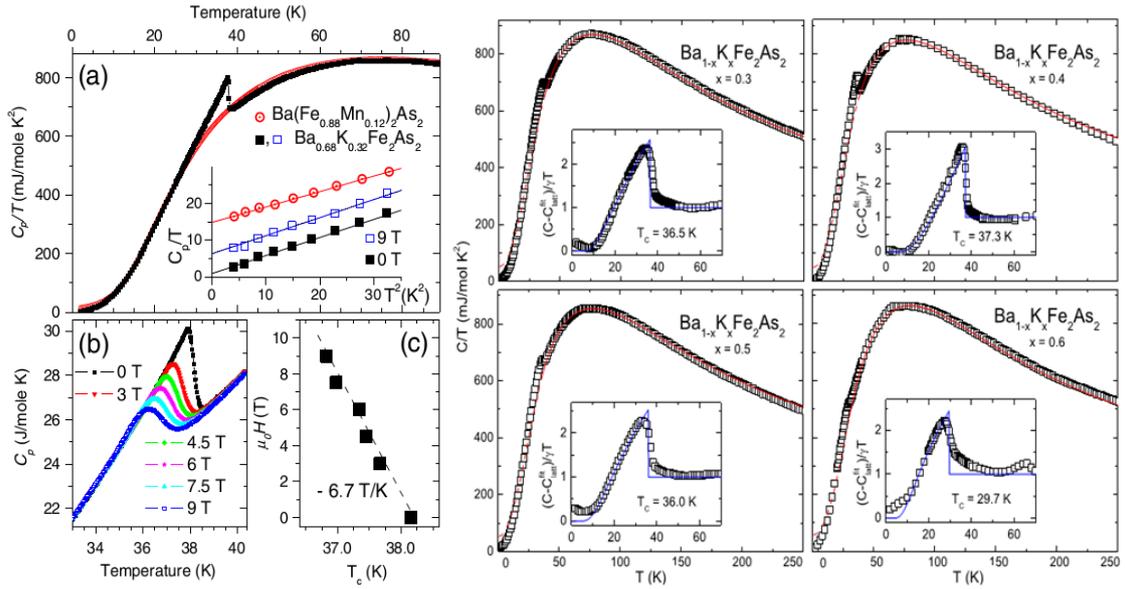

**Figure 14 (left)**(a) *Temperature dependence of the specific heat $C_p/T$ of $Ba_{0.68}K_{0.32}Fe_2As_2$ (squares) and $Ba(Fe_{0.88}Mn_{0.12})_2As_2$(circles). The inset shows a plot of $C_p/T$ vs $T^2$ at low T for both samples. The lines represent the best fit to $C_p(T)/T = \gamma(0) + \beta T^2$. (b,c) Temperature dependence of $C_p$ of $Ba_{0.68}K_{0.32}Fe_2As_2$ near $T_c$ measured at different external magnetic fields applied perpendicular to the ab plane [36].(**Right**) C/T as a function of T of $Ba_{1-x}K_xFe_2As_2$ for x=0.3, 0.4, 0.5, 0.6. Dashed (red) line is the extrapolation of the fit in lower temperature. Insets describe the difference between C/T and modelled phonon contribution (normalised by the normal-state Sommerfeld coefficient γ )[37].*

of Figure 14. Higher $T_c$s and sample quality issues make it difficult to estimate $\gamma_n$ for most 1111 materials. Low $T_c$ compounds, such as $FeSe_{0.88}$, ($\gamma_n = 9.2$ mJ/moleK$^2$) [37] $T_c \approx 8$K, LaFePO, ($\gamma_n = 7$ mJ/moleK$^2$) [38], $T_c \approx 5$-6 K, and $KFe_2As_2$, ($\gamma_n = 69$ mJ/moleK$^2$ and RRR=67) [39], $T_c$=3.4 K, have $\gamma_n$s that are more easily determined. There are evidences on the sample dependence of $\gamma_n$ in FePn/Ch compounds. $\gamma_n = 69$mJ/moleK$^2$ for sample with RRR=67 of polycrystalline $KFe_2As_2$[40]. Hashimoto et al., referenced an unpublished result for $\gamma_n$ of 93 mJ/moleK$^2$ with RRR>1200 and for a single crystal $KFe_2As_2$ with RRR=650, $\gamma_n$ = 102 mJ/moleK$^2$ [41]. Table-III & Table-IV shown below provide estimations of $\gamma_n$ and reached conclusions of various authors of the references indicated.

**Table-III**: $\gamma_n$ for $Ba_{1-x}K_xFe_2As_2$

| X | $T_c$ in(K) | $\gamma_n$ mJ/mole-K$^2$ | Ref. | Conclusion |
|---|---|---|---|---|
| 0-0.6 | 23K | 50 – 65 | [37] | Simple BCS, single band picture |
| 0.32 | 38.5K | 50 | [36] | Strong coupling, two band model |
| 0.4 | – | 63 | [42] | Multi-gap effect, complex FS |

**Table-IV:** Specific heat $\gamma_n$ and $T_c$ for unannealed and annealed* BaFe$_{2-x}$Co$_x$As$_2$

| x | $T_c$ in (K) | $\gamma_n$(mJ/moleK$^2$) | Reference |
|---|---|---|---|
| 0.08/0.09 | 5.8/5.6,8.0* | 14.9/13.7,14* | [43] [44] [45] |
| 0.10 | 19.5 | 17.2 | [43] |
| 0.11 | 21.5 | 19 | [43] |
| 0.115 | 24.3 | 21.3 | [43] |
| 0.15/0.16 | 22.1/18,22* | 22.1/18,22* | [43] [44] [45] |
| 0.18 | 20 | 20 | [43] |
| 0.22/0.21 | 17/23,20* | 17/23.2,20* | [43] [44] [45] |
| 0.24 | 14.6 | 14.6 | [43] |
| 0.31 | 16 | 16 | [43] |

Comparison of these measured values of $\gamma_n$ with those obtained from band structure calculations may provide information about DOS near Fermi surface and may provide partial clue on mechanism of superconductivity etc. Measured low temperature $\gamma_n$ is lower than the calculated values. Due to magnetic order/SDW transitions in the 21311, 122*, 1111, 122 parent compounds there is depletion of DOS at the FS which may be the reason for such discrepancies. Thus, the band structure calculations may be compared with experimental values of $\gamma_n$ only for a non-magnetic doped system or for a non-magnetic 111 or 11 compound. $\gamma_n = \frac{1}{3}\pi^2 k_B^2 N(0)(1 + \lambda)$ which implies, $N(0)(1 + \lambda) = \frac{0.42\gamma_n}{n}$ where $n = 5$ for 122 compound (a mole of 122 contains 5 atoms). Therefore, in principle by obtaining N(0) from band structure calculation accurately, one can find the value of effective coupling constant $\lambda$. For example, using $n = 2$, $\lambda \sim 1$ when calculated from the data [38]. On the other hand, following the data of Kant *et al.*, [37], $\gamma_n = 54 mJ/moleK^2$ and $n = 5$, leads to a $\lambda \sim 3.1$ which is very large perhaps indicative of the fact that $N(0)$ has been underestimated, for the reason mentioned above. If $(1+ \lambda)$ (known as the mass renormalization) were solely due to electron-phonon interaction (for $FeSe_{1-x}$, $LaFePO$) it require $\lambda_{el-ph} \sim 0.8$ and $0.6$ [38][46] respectively. In fact it is consistent with the calculated values of $(1+\lambda) \sim 2.05/1.7$ [38]. However from calculation of Subedi *et al.*, (2008) ($\lambda_{el-ph} = 0.17$ for FeSe) [47] and Boeri *et al.* (2008) ($\lambda_{el-ph} = 0.21$ for LaFeAsO) [48] it is clear that even these low $T_c$ FeAS/Ch are not conventional BCS type electron-phonon pairing superconductors. Spin fluctuation, electron-electron interactions possibly others are important in these cases.

The specific heat jump at $T_c$, $\Delta C/T_c$ is a well-known parameter, historically, to be compared with other conventional as well as other classes of materials. Bud'ko, Ni & Canfield (2009) (BNC) [49] provides such a data for 14 different doped samples of 122 compounds that scales $\Delta C/T_c = aT_c^2$ with $a \sim 0.056 mJ/moleK^4$ for all those 122 superconductors. Based on the work of J. S. Kim *et al.*, a modified BNC plot may be drawn that includes other FePn/Ch superconductors other than 122 only along with conventional electron-phonon superconductors (elements with $T_c$>1K and A-15 superconductors). Several unconventional heavy fermion superconductors are also included [50]. The modified BNC plot with $\frac{\Delta C}{T_c} \approx T_c^{1.89}$ and $a = 0.083$ mJ mol$^{-1}$K$^{-1}$ is presented in figure 15. Both the conventional and heavy fermion superconductors show different behaviour

($\frac{\Delta C}{T_c} \approx T_c^{0.9}$) than that of the FePn/Ch superconductors. This unconventional behaviour of FePn/Ch superconductors compared to other classes of superconductors clearly indicates that the mechanism of superconductivity is different in these materials may be different.

There are different source of errors in the determination of intrinsic $\Delta C/T_c$. Depending on the sample quality transitions can be quite broadened in temperature to determine the $T_c$ accurately. One can analyse such broadened transitions by so-called "equal area construction"; see the right side figure of Fig. 15. In this method, the low temperature superconducting state data up to the initial bend over in C/T at $T_c^{low}$ are extrapolated linearly further as $C_{sc}^{ex}/T$; similarly, the normal state data are extrapolated linearly as $C_n^{ex}/T$ to lower temperature. Then ΔC is constructed at a temperature around midway between $T_c^{onset}$ and $T_c^{low}$ at $T_c^{mid}$ with the area (entropy) between the linearly extrapolated $C_{sc}^{ex}/T$ and the actual measured data below $T_c^{mid}$ equal to the area (entropy) between the measured data above $T_c^{mid}$ and the extrapolated $C_n^{ex}/T$ from above $T_c^{onset}$. This gives the correct

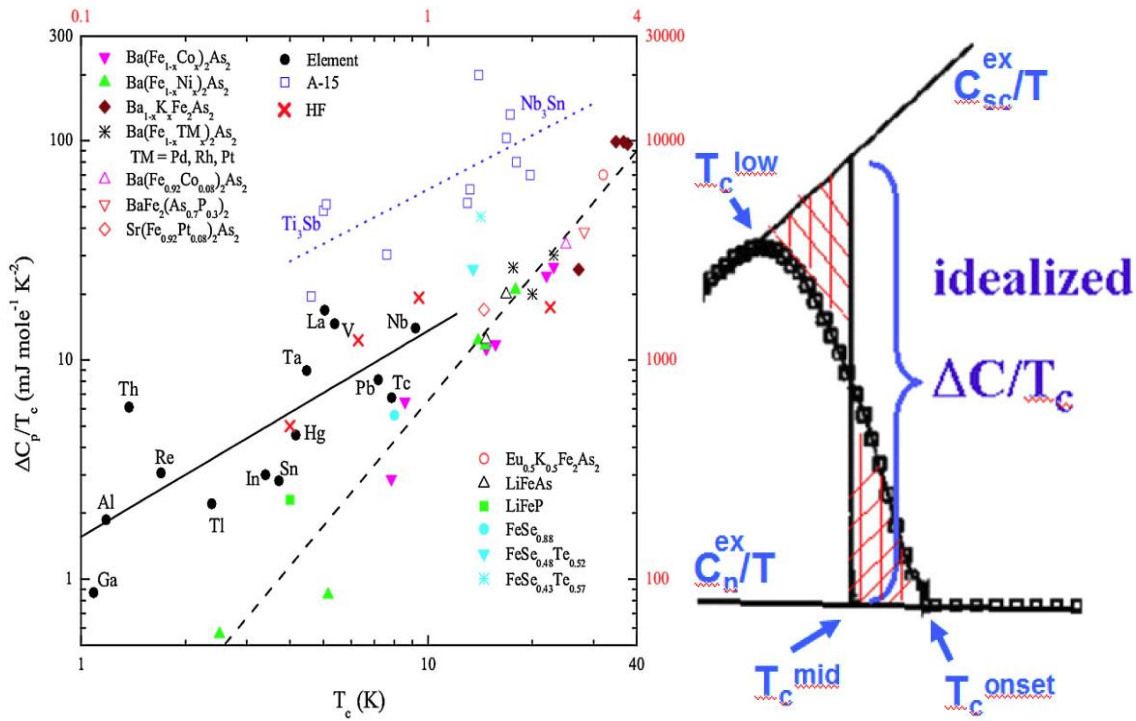

**Figure15**. *Expanded BNC plot based on the work by J. S. Kim et al. 2011*(**left**)*[50].Equal are construction method for calculating $\Delta C/T_c$ in a broadened transition squares indicate data points and crosshatching are equal area (entropy)*(**right**).

estimation of the superconducting state entropy at $T_c$. There are many samples of the FePn/Ch superconductors which have finite $\gamma$ in the superconducting state that is likely not intrinsic.

Other variants of similar kinds of studies are presented in Fig. 16. The left side figure is a log-log plot of $\frac{\Delta C}{T_c}$ vs $T_c$ showing $\frac{\Delta C}{T_c} \propto T_c^2$ whereas the right hand side figure of Fig. 16 is a simple plot of $\frac{\Delta C}{T_c}$ vs $T_c$ for various Fe-based superconductors. The slope of the curve clearly indicates $\frac{\Delta C}{T_c} \propto T_c^2$.

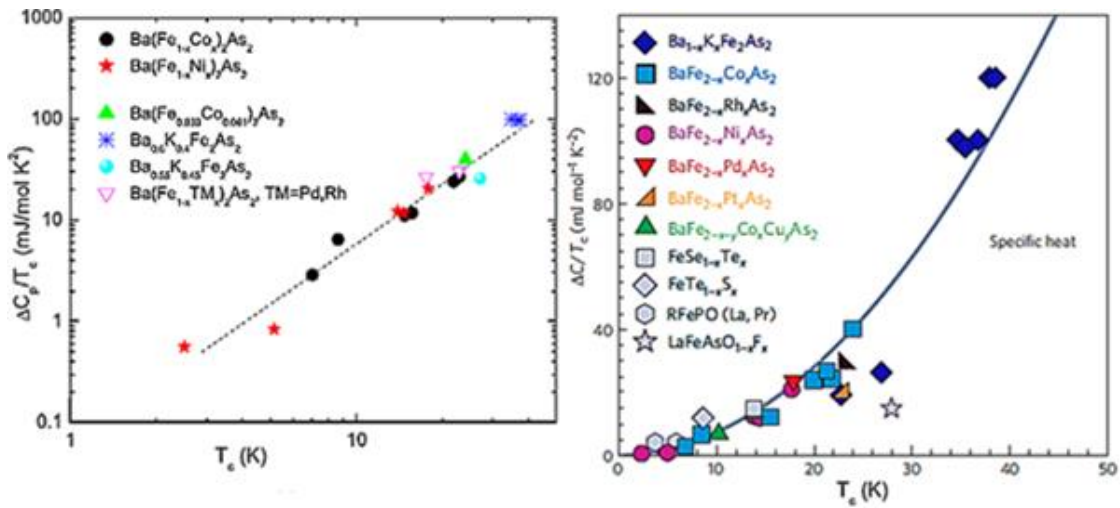

**Figure 16**. *A log-log plot showing $\Delta C/T_c \propto T_c^2$ for doped BaFe₂As₂ (left)*[51]. $\frac{\Delta C}{T_c}$ *vs. $T_c$ plot of different Iron based superconductors(right)* [52].

**Fermi-ology**

An atomically flat sample is illuminated by a beam of monochromatic light in angle resolved photoemission (ARPES) experiments. Due to the photoelectric effect, the sample emits electrons. The kinetic energy and direction of these electrons are measured by the apparatus. The flat surface of the sample has translational symmetry. Therefore, as electrons escape from the solid, linear momentum is conserved parallel to the surface. The photon momentum is smaller than that of electrons by 2 orders of magnitude and can be neglected. ARPES thus directly measures the components of electron momentum that are parallel to the surface. As a result, ARPES is almost an ideal tool for imaging the Fermi surface of solids (especially low dimensional systems). In this section we shall reproduce a glimpse of ARPES results available in literature and its possible impacts.

Ding *et al.*, observed two superconducting gaps with different values: a large gap ($\Delta \sim 12$ meV) on the two small hole-like and electron-like Fermi surface (FS) sheets, and a small gap ($\sim 6$ meV) on the large hole-like Fermi surface for 122 compound ($Ba_{0.6}K_{0.4}Fe_2As_2$ at 15 K), Fig.17 [53]. Both the gaps, terminating simultaneously at the bulk transition temperature ($T_c$), are node less and nearly isotropic around their respective Fermi surface sheets. Similar Fermi-ology is also applicable to 1111 compounds. However, there are some compounds that have only either hole like FS or electron-like FS but not both, such examples are produced in Fig. 18. In general evaluation of the FS with doping electron and/or hole is quite complicated and variation of $T_c$ with electron or hole doping is not completely symmetric. Fermi-ology of FePn/Ch superconductors plays a major role in the pairing mechanism, symmetry and structure of the superconducting energy gap.

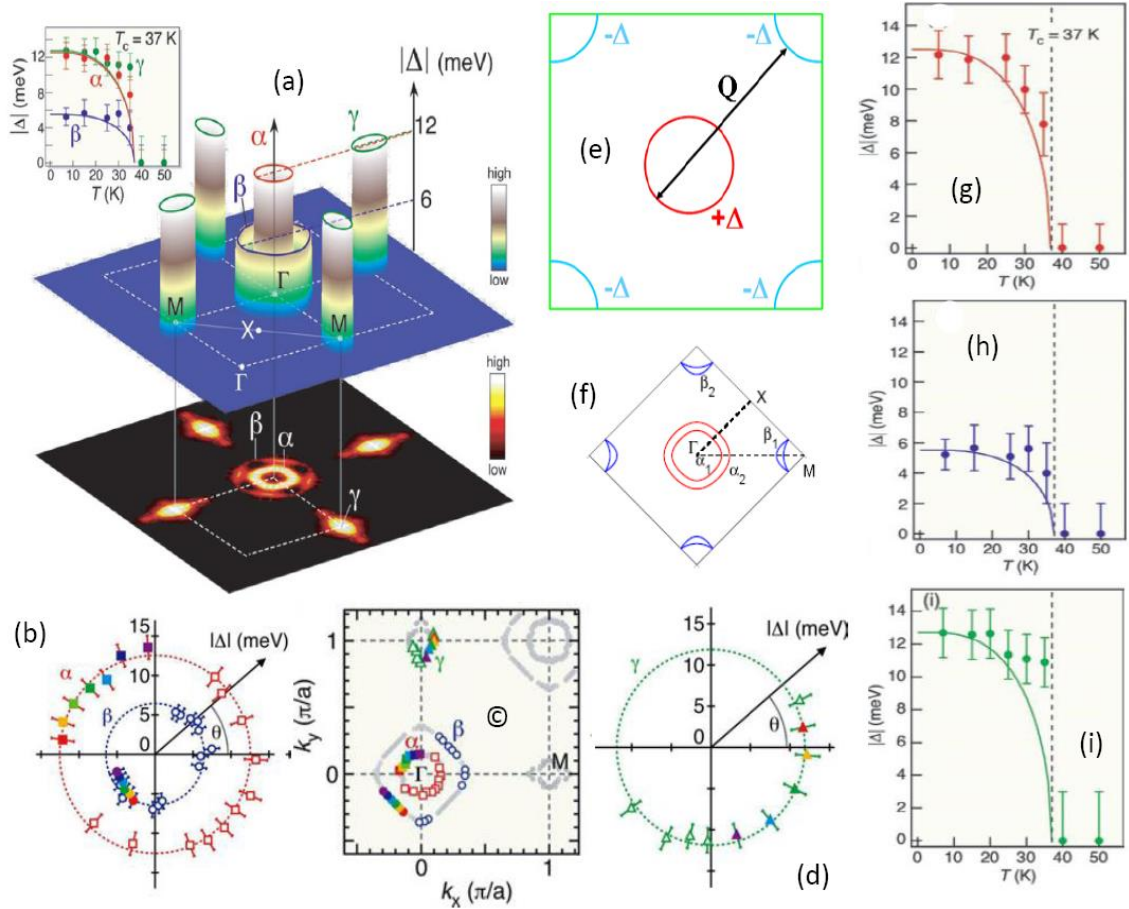

**Figure** 17. *ARPES image of Fermi surface (FS) showing Fermi-ology of Fe-based superconductors for 122 systems. There are two hole-like concentric pockets $\alpha, \beta$ around the $\Gamma$-point (a, c) and electron-like Fermi-pockets around M-points. The outer hole like Fermi pockets have smaller gap (b, h) whereas the inner hole like $\alpha$ and electron-like $\gamma$ –Fermi pockets have same but larger gap magnitudes (b, d, g, i). (e, f) Schematic two dimensional FS in full and folded magnetic Brillouin zone respectively.*

There are theoretical predictions and experimental evidences of $s_\pm$ state [54] and prediction of other pairing mechanism like $s_{++}$ state mediated by orbital fluctuations [55, 56] or coexistence of both. Here $s_\pm$ and $s_{++}$ pairing symmetry implies $l = 0$ angular momentum state of the cooper pair wave function which changes sign and does not change sign from one sheet of the FS to other respectively. If the pairing is due to exchange of spin fluctuations then even if the pairing coupling interaction turns to be repulsive it can still lead to attractive pairing if the excitation is being exchanged between parts of the Fermi surface with opposite signs of the order parameter (see Fig. 17 e). Observed coexistence of SDW which occurs due to FS nesting and superconductivity in (122) $BaFe_{2-x}Co_xAs_2$ $s_\pm$ pairing symmetry is a very strong possibility. Detailed discussions on pairing symmetry and related mechanisms are beyond the scope of this review (see for example [57]). 122* materials have large local moment [58] and with no hole pocket [59]. There are also materials like $KFe_2As_2$ having very small or no electron like Fermi surface present (Fig. 18) [60] exactly contrary to the FS topology of 122* materials [61]. These materials thus would favour $s_{++}$ pairing symmetry (issue of pairing symmetry is far from being settled).

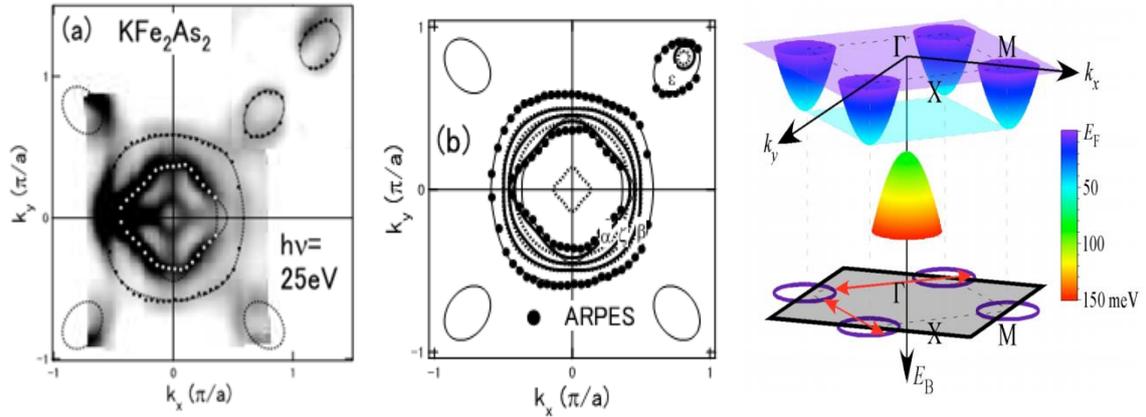

**Figure**18. *Fermi surface of $KFe_2As_2$ obtained by ARPES shows absence of electron like Fermi surface (reproduced from [60]). Absence of hole-like Fermi surface in superconducting $K_{0.8}Fe_{1.7}Se_2$ revealed by ARPES [61].*

## $2\Delta/k_BT_c$ ratio:

$2\Delta/k_BT_c$ ratio represents the fundamental coupling strength of superconductivity. $2\Delta/k_BT_c$ ratios of Fe-based superconductors are summarized by Evtushinsky *et al*., [62]. Although there are evidences that many of the Fe-based materials are multi-band materials, there exist two kinds of gaps one larger gap and the other small gap. In hole doped $Ba_{1-x}K_xFe_2As_2$ the large gap opens on the inner Γ-barrel and the propeller-like structure around the X point, while the small gap opens only on the outer Γ-barrel [53, 62, 63]. It is noted from a recent ARPES studies of the electron-doped compound $BaFe_{1.85}Co_{0.15}As_2$ that the smaller gap opens on the bands in the vicinity of X point, while the large one opens on the bands around Γ [64]. $2\Delta/k_BT_c$ ratio of Fe-based superconductors extracted from various experimental studies is presented in Fig. 17. First we present the expansion of abbreviations of various experimental studies used in the x-axis of Fig. 17. PCAR abbreviated for Andreev reflection spectroscopy [65-72], ARPES [68,53,73-76,64], SI as surface impedance [82], critical magnetic field ($H_{c1}$) [77], muon spin rotation (μSR) [78-81], scanning tunnelling spectroscopy (STS) [85-88]. It is evident that the gap is small ($2\Delta_{small}/k_BT_c < 3$) on the outer Γ-barrel, and large on the other parts of the Fermi surface ($2\Delta_{large}/k_BT_c \sim 7.0$). This makes these compounds very different from other classes of superconductors.

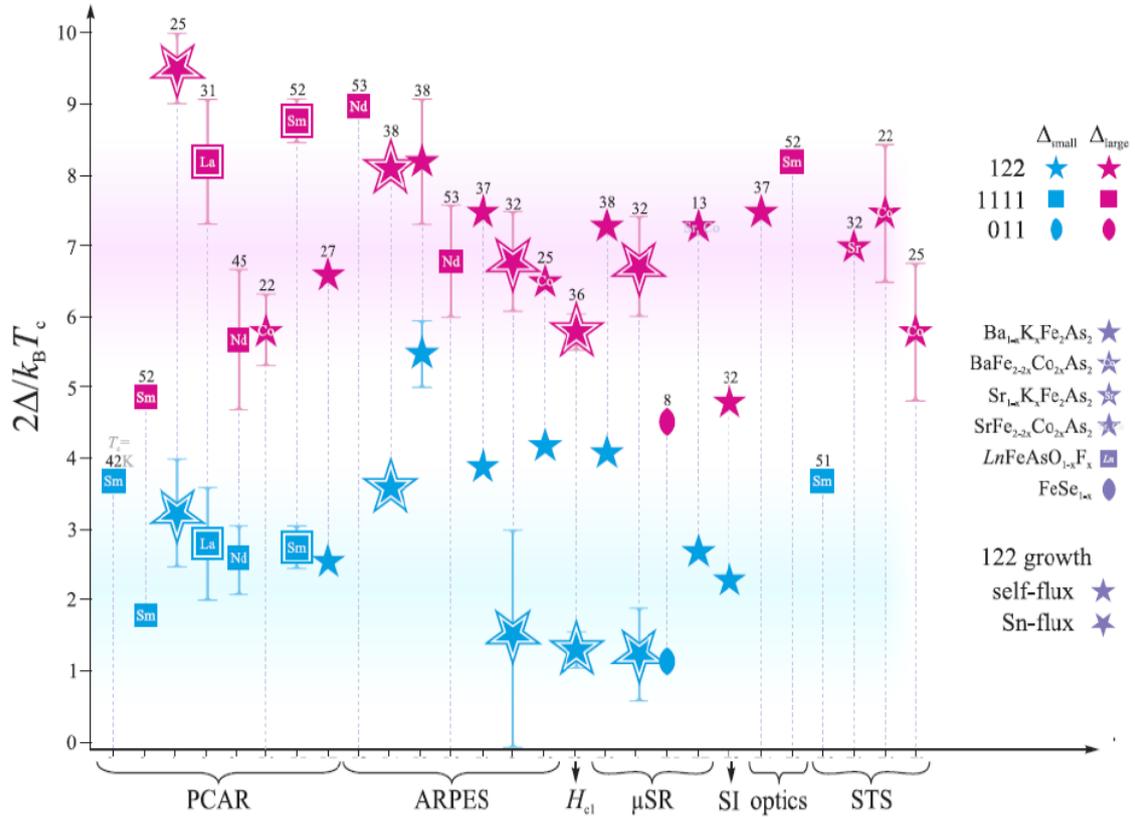

**Figure 17**. *$2\Delta/k_BT_c$ ratio of various Fe-based superconductors as concluded from various experimental studies, taken from ref. [62]. See text for expansion of the abbreviations and symbol explanations. It is evident that the gap is small ($2\Delta_{small}/k_BT_c < 3$) on the outer Γ-barrel, and large on the other parts of the Fermi surface ($2\Delta_{large}/k_BT_c \sim 7.0$). For sources of experimental data see text.*

In conventional superconductors, the electron-phonon interaction responsible for electron pairing and superconductivity was established by tunnelling and neutron scattering experiments, where dips in the second derivative of the tunneling current correspond to phonon modes observed by inelastic neutron scattering experiments [89–91]. To obtain the equivalent information for high-$T_c$ Fe-based superconductors, it is important to identify the electron-boson coupling [92] and its connections to superconductivity. For bosonic ''pairing glue'' mediated superconductors, the ''glue'' may arise from the usual electron-phonon interactions or the exchange of particle-hole spin fluctuations characterized by the imaginary part of the dynamic susceptibility, which is seen in inelastic magnetic neutron scattering measurements [93-96]. In one of the leading theories, superconductivity arises from quasi-particle excitations between the electron and hole pockets near M and Γ points of the Brillouin zone, respectively. One of the consequences of opening up electronic gaps in the superconducting state is that there should be a neutron spin resonance. The energy of the resonance should be coupled to the addition of the hole and electron superconducting gap energies ($\hbar\omega = |\Delta_{k+Q}| + |\Delta_k|$), and the intensity of the mode should follow the superconducting order parameter [97,98]. Below is a table containing various Fe-based compounds, corresponding $T_c$ in K, the resonance energy obtained from the resonance peak in inelastic neutron scattering experiments, the scaling of resonance energy with $k_B T_c$ and references from which the data is obtained.

| Compound | $T_c$(K) | $\Omega_{res}$(meV) | $\Omega_{res}/k_B T_c$ | Reference |
|---|---|---|---|---|
| $BaFe_{2-x}Co_xAs_2$, $x = 0.08$ | 11 | 4.5 | 4.9 | [99] |
| $x = 0.094$ | 17 | ~4.5 | 3.2 | [100] |
| $x = 0.13$ | 23 | ~10 | 5.2 | [101] |
| $x = 0.148$ | 22.2 | 8.3 | 4.5 | [102] |
| $x = 0.15$ | 25 | 9.5 | 4.6 | [103] |
|  | 25 | 9.6, 10.5a | 4.6, 5.0 | [104] |
| $x = 0.16$ | 22 | 8.6 | 4.7 | [105] |
| $BaFe_{2-x}Ni_xAs_2$ $x = 0.075$ | 12 | 5, 7a | 5.0, 7.0 | [106] |
| $x = 0.09$ | 18 | 6.5, 8.8a | 4.3, 5.9 | [107] |
| $x = 0.1$ | 20 | 7.0, 9.1a | 4.2, 5.5 | [108] |
| $x = 0.15$ | 14 | 6, 8a | 5.1, 6.9 | [106] |
| $FeSe_{0.4}Te_{0.6}$ | 14/14.6 | 6.5/7.1 | 5.6 | [109] |
| $FeSe_{0.5}Te_{0.5}$ | 14 | 6/6.5 | ~5.6 | [110] |
| $LaFeAsO_{1-x}F_x$ $x = 0.057/0.082$ | 25/29 | 11 | 5.3/4.6 | [111] |
| $Ba_{0.6}K_{0.4}Fe_2As_2$ | 38 | 14 | 4.4 | [112] |
| $BaFe_2(As_{0.65}P_{0.35})_2$ | 30 | 12 | 4.8 | [113] |
| LiFeAs | 17 | 8 | ~5–6 | [114] |

It is evident that the $\Omega_{res}$ scales with the corresponding $T_c$ which may thus signify spin fluctuation mechanism of superconductivity in Fe-based materials.

**Summary**

This is a transcript of the invited talk delivered at the Advances in Material Sciences and Technology, November 2012 held at the Kakatiya University, India. Materials used in this review have already been published in different journals and reviews [115]. This no way represents a complete story of this still very fast emerging field but merely part of the directions covered in the talk. In case any of the data presented in this review are not cited with reference is merely accidental and do not intend to represent authors personal data.

Fe-based superconductors are fundamentally *different* from other classes of superconductors in a number of ways; (a) coupling of magnetism, superconductivity and orbital degrees of freedom --- possible connection of structural transition to orbital degree than lattice. This is evidenced by presenting a number of phase diagrams in the PHASE DIAGRAM section. In a number of cases structural and magnetic transition temperatures are identical, a unique feature to these systems. (b) The specific heat jump scales very differently than other conventional as well as unconventional superconductors. This feature is established in detail in the SPECIFIC HEAT section. (c) They have very different Fermi-ology made of Fe-*d* electrons than other classes of superconductors which changes drastically with electron or hole doping as discussed in detail in the Fermi-ology sub-section. (d) Although numerous theoretical/experimental data suggest to multiband Fermi surface and multi-gap, there are strong evidences from various experimental studies of two –gaps (i) one large and (ii) small gaps giving rise to two sets of $2\Delta/k_B T_c$ ratios, average about 7 and 3 respectively (see Fig.17). (d) Unlike conventional superconductors Fe-based superconductors do not exhibit any *O* istope effect but Fe istope effect instead. (e) The spin susceptibility in a large number of compounds produces linear temperature dependencies. (f) Spin resonance energy which is also observed in high $T_c$ cuprates, scales with $T_c$ linearly. Finally, there are many exotic theories, which are kept outside the purview of this review --- mere facts are accumulated. This review in no way covers all the aspects of this class of superconductors and many important works of many authors are left out. The review includes a large number of references; we hope will provide sufficient platform for beginners.